\documentclass{iaus}
\usepackage{graphicx}

\def\MH{\mathrm{[M/H]}}
\def\teff{\mathrm{T_{eff}}}
\def\kms{\mbox{km~s$^{-1}$}}
\def\logg{\log g}
\def\ltsima{$\; \buildrel < \over \sim \;$}
\def\simlt{\lower.5ex\hbox{\ltsima}}
\def\gtsima{$\; \buildrel > \over \sim \;$}
\def\simgt{\lower.5ex\hbox{\gtsima}}

\title[Present state and promises of the RAVE survey] 
{Present state and promises to unravel the structure and 
kinematics of the Milky Way with the RAVE survey}

\author[Steinmetz, Siebert, Zwitter]   
{M. Steinmetz$^1$, A. Siebert$^2$, T. Zwitter$^3$ and the RAVE collaboration
}

\affiliation{$^1$Astrophysikalisches Institut Potsdam, Potsdam, Germany
\\[\affilskip]
$^2$Observatoire de Strasbourg, Strasbourg, France
\\[\affilskip]
$^3$
University of Ljubljana, Faculty of Mathematics and Physics, Ljubljana, Slovenia
}

\pubyear{2008}
\volume{254}  
\pagerange{1--6}
\jname{The galaxy disk in cosmological context}
\editors{A.C. Editor, B.D. Editor \& C.E. Editor, eds.}
\begin{document}

\maketitle

\begin{abstract}
The RAdial Velocity Experiment (RAVE) is an ambitious survey to measure the
radial velocities, temperatures, surface gravities, metallicities and
abundance ratios for up to a million stars using the 1.2-m UK Schmidt
Telescope of the Anglo-Australian Observatory (AAO), over the period 
2003 -- 2011. The survey represents a big advance in our understanding of 
our own Milky Way galaxy. The main data product will be a southern hemisphere
survey of about a million stars. Their selection is based exclusively on
their $I$--band colour, so avoiding any colour-induced bias. RAVE is expected
to be the largest spectroscopic survey of the Solar neighbourhood in the
coming decade, but with a significant fraction of giant stars reaching out to
10 kpc from the Sun. RAVE offers the first truly representative inventory of
stellar radial velocities for all major components of the Galaxy. Here we
present the first scientific results of this survey as well as its second
data release which doubles the number of previously released radial
velocities. For the first time, the release also provides atmospheric
parameters for a large fraction of the second year data, making it an
unprecedented tool to study the formation of the Milky Way. Plans for further
data releases are outlined. 
\end{abstract}

\firstsection 
\section{Introduction}
                                                                
It is now widely accepted that the Milky Way galaxy is a suitable laboratory
to study the formation and evolution of galaxies. Despite the fact that the
Galaxy is one unique system, understanding its formation holds important keys
to study the broader context of disc galaxy formation. Thanks to the past and
ongoing large surveys such as Hipparcos, SDSS, 2MASS or DENIS, we have access
to data which allow us to refine our knowledge of Galaxy formation. However,
with the exception of the SDSS survey, which mainly samples the halo of the
Galaxy, the full description of the 6D phase space, i.e.\  the combination of
the position and velocity spaces, is not available due to the missing radial
velocity and/or distance.

With the advent of multi-fiber spectroscopy, combined to the large field of
view of Schmidt telescopes, it is now possible to acquire in a reasonable 
amount of time spectra for a large sample of stars that is representative for
the different populations of the Galaxy. Spectroscopy enables us to
measure the generally missing radial velocity, which in turn allows us to
study the details of Galactic dynamics. Spectroscopy also permits to measure
the abundance of chemical elements in a stellar atmosphere which holds
important clues on the initial chemical composition and its subsequent metal
enrichment.  The measurement of the radial velocity and of the chemical
abundances as well as the derivation of stellar temperature and gravity in
order to complement existing catalogues is the main purpose of the RAVE
project.

RAVE is using the 6dF multi-fiber spectroscopic facility at the UK
Schmidt telescope of the Anglo-Australian Observatory in Siding Spring,
Australia. The 6dF enables us to collect up to 150 spectra in one single
pointing, with an average resolution of 7500 in the Calcium triplet region
around 8500~\AA. This medium resolution allows the measurement of
accurate radial velocities ($\sim 2$ \kms) as well as atmospheric parameters
($\teff$ , $\logg$, $\MH$) and chemical abundances. In Section 2 we present
the first scientific outcomes obtained using the RAVE data, while Section 3
presents the second data release of the RAVE project (DR2) and discusses the
current status and prospects of the project.

\section{First Results from the Survey}

In this section we outline some of the recently published results based on
RAVE data. 

\subsection{The Escape Velocity of the Galaxy}

In \cite{smith_2007} we revisited the local escape speed of our Galaxy by
combining a sample of high velocity stars detected by the RAVE survey with
previously known ones. Using a maximum likelihood technique, we find 
$498 < v_{esc} < 608$~\kms\ at the 90\%\ confidence level, with a median
likelihood of 544~\kms. This result demonstrates the presence of a dark halo
in the Milky Way, but simultaneously argues for a halo of relatively low
circular velocity ($v_c \approx 140$~\kms).

\subsection{Streams (or lack thereof) in the Solar Neighborhood}

In \cite{seabroke_2008} we searched the CORAVEL and RAVE survey for
signatures of vertically infalling stellar streams in the Solar vicinity.
Using a Kuiper test, we demonstrated that the Solar neighborhood is empty of
any vertical streams containing more than a few hundreds of stars.
Therefore, we confirm recent simulations that are favoring a model in which
the Sagittarius stream is not entering the Solar neighborhood. We also argue
against the Virgo overdensity crossing the disc near the Sun.

\subsection{The Vertical Structure of the Galactic Disc}

Combining RAVE data with photometric and astrometric catalogues, in
\cite{veltz_2008} we were using G-- and K--type stars towards the Galactic
poles in order to identify whether there is a kinematic discontinuity
between the thin and thick discs. We conclude that such a discontinuity
indeed exists, which is a strong constraint on the formation scenario of the
thick disc: it is arguing against continuous processes such as scattering by
spiral stellar or molecular arms, but favoring violent processes such as the
accretion of or violent heating by a satellite.

\begin{figure}[b]
\begin{center} \includegraphics[width=5.0in]{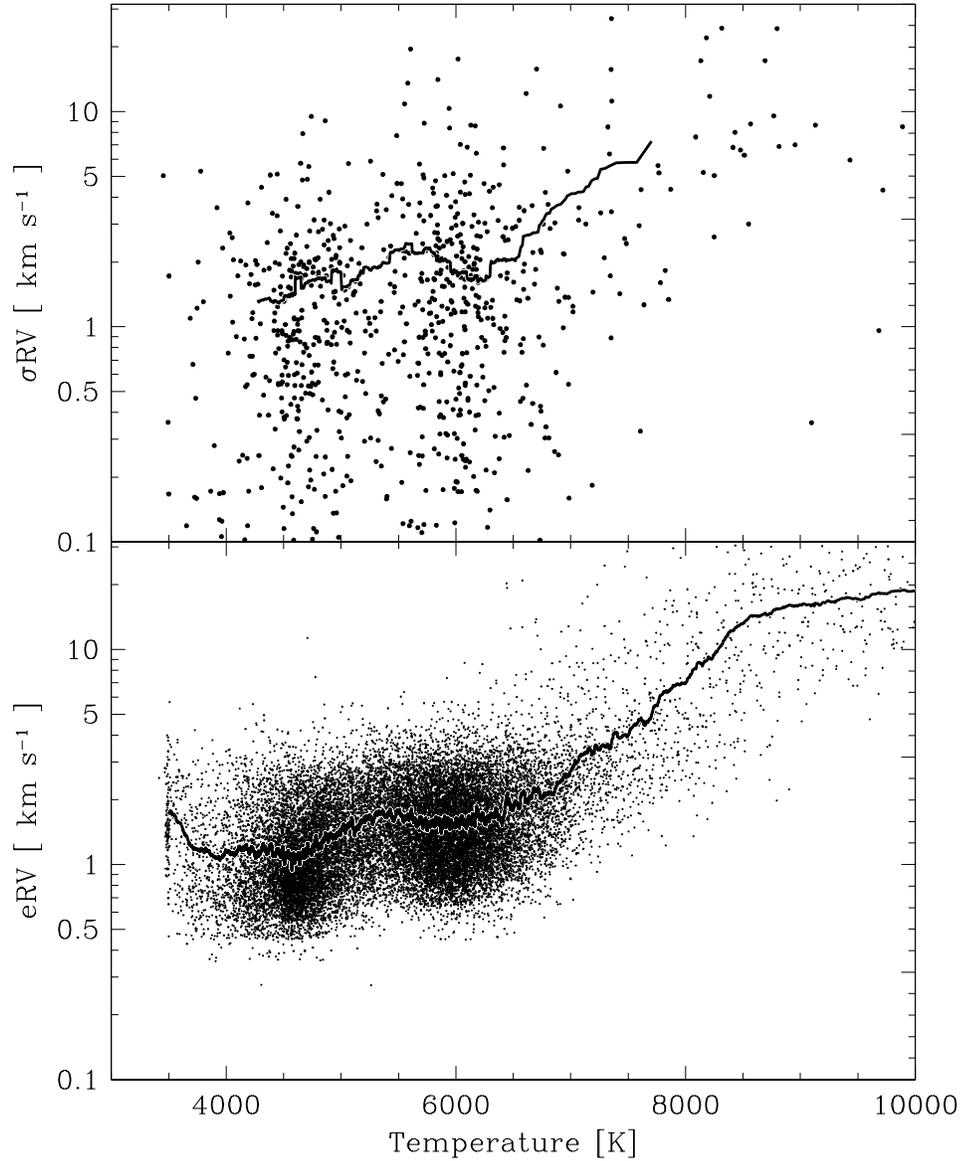} 
\caption{Radial velocity errors for stars in the second RAVE data release as
a function of stellar temperature. The error is the internal velocity error
(bottom) and the standard deviation of velocities determined from repeated
observations of the same object (top). Lines trace a smoothed average
dependence, with a boxcar width of 100 points.  }   \label{fig1}
\end{center}
\end{figure}                                                                
                                                                 \subsection{The Tilt of the Velocity Ellipsoid}
                                                     
Using RAVE red clump giants towards the South Galactic pole, in
\cite{siebert_2008} we have measured the inclination of the velocity
ellipsoid at 1 kpc below the Galactic plane. The value of the tilt, $7.3 \pm
1.8^o$, is consistent with either a short scale length for the disc 
($R_d \sim 2$~ kpc) if the halo is oblate, or a long scale length 
($R_d \sim 3$~kpc) if the halo is prolate. Combined to independent
measurements of the minor-to-major axis ratio of the halo, which prefers an
almost spherical halo. A scale length of the disc in the range 
[2.5 -- 2.7] kpc is preferred.
                                                                             \subsection{Diffuse Interstellar Bands}                                       
                                                                             In \cite{munari_2008} we used spectra of hot stars from the RAVE survey to
investigate the properties of 5 diffuse interstellar bands (DIB) in the Ca
triplet region. Our findings indicate that the DIB at 8620.4~\AA\ is 
strongly correlated to reddening and follows the relation 
$E(B-V ) = (2.72\pm 0.03)\times$EW, where EW is the equivalent width of the
DIB in \AA. This DIB is thus a suitable tracer of general Galactic reddening
in stellar spectra. On the other hand the existence of the DIB at 8648~\AA\
is confirmed, but its intensity or equivalent width does not appear to
correlate with reddening.
                                                                \section{Second data release: Current Status and Ongoing efforts}

\begin{figure}[b]
\begin{center} \includegraphics[width=3.4in,angle=270]{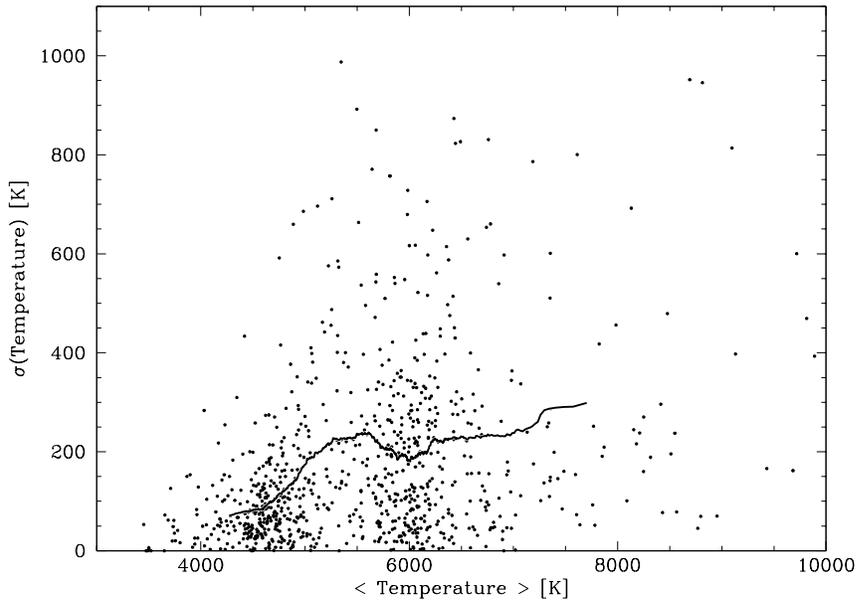} 
\caption{Temperature accuracy as judged from repeated observations of the
same object in the second data release. Standard deviation of temperature is
given as a function of its average value. The line is a smoothed average
using the adjacent 50 cooler and 50 hotter objects.  }   \label{fig2}
\end{center}
\end{figure}

\begin{figure}[b]
\begin{center} 
\includegraphics[width=3.4in,angle=270]{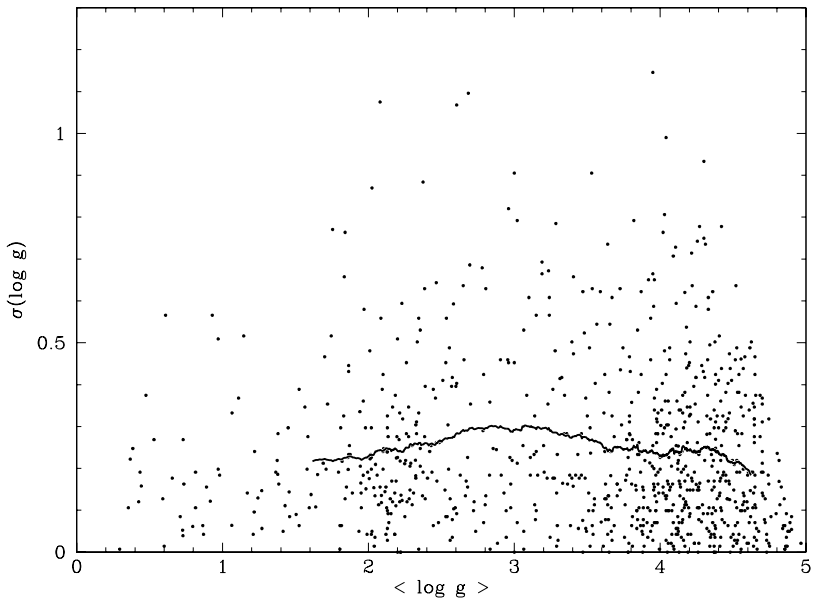} 
\caption{Accuracy of surface gravity as judged from repeated observations.
Plot follows the style of Figure~\ref{fig2}.  }   \label{fig3}
\end{center}
\end{figure}

\begin{figure}[b]
\begin{center} \includegraphics[width=3.4in,angle=270]{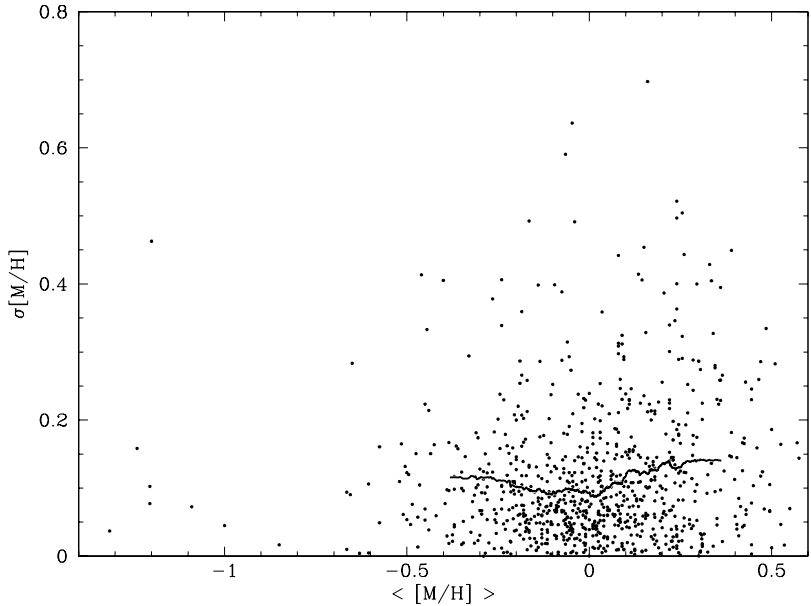} 
\caption{Accuracy of metallicity as judged from repeated observations. Plot
follows the style of Figure~\ref{fig2}.   }   \label{fig4}
\end{center}
\end{figure}

\begin{figure}[b]
\begin{center} \includegraphics[width=5.5in]{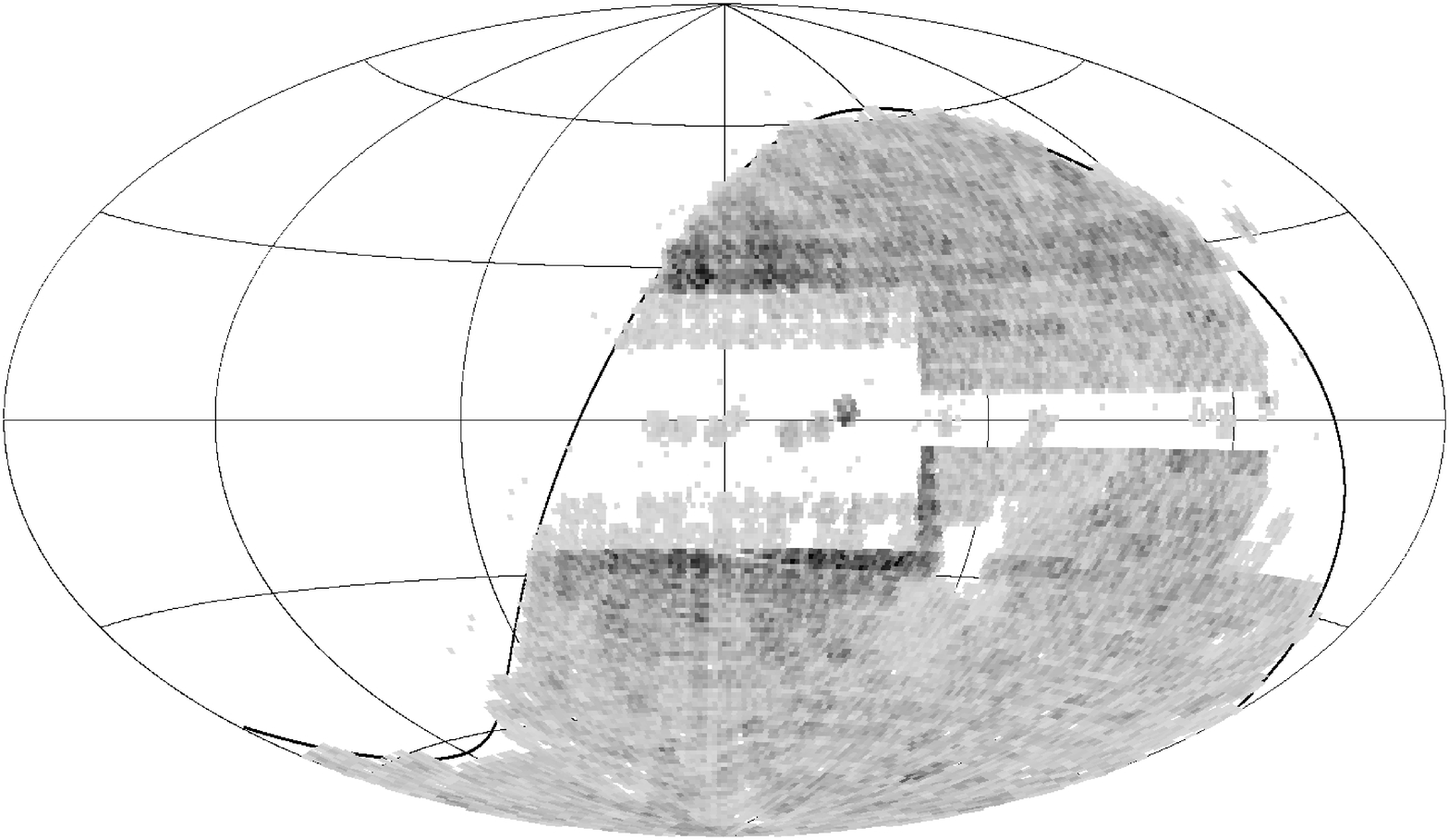} 
\caption{Aitoff projection in Galactic coordinates of the number of RAVE
spectra per square degree. Only the spectra observed before 6th of July 2008
and which successfully passed the quality controls are included. They cover
16,730 square degrees of the Sourhern sky, with a mode of 8 spectra per
square degree. Density is linearly coded in shades of grey from  zero in
white to the maximum of 93 spectra per square degree in black. Spectra on
the Galactic plane were observed for calibration and test purposes. The
S-shaped curve marks the celestial equator. 
 }   \label{fig5}
  \end{center}
   \end{figure}

                                                                             \subsection{Second Data Release}
                                                                             In July 2008, the RAVE collaboration has released the 2nd catalogue of
reduced RAVE data (DR2) \cite[(Zwitter et al. 2008)]{zwitter_2008}.  DR2
covers an area of $\sim 7200\, \mathrm{deg}^2$ in the southern hemisphere at
galactic latitudes larger than $25^\mathrm{o}$. As for the first data
release (DR1) \cite[(Steinmetz et al. 2006)]{steinmetz_2006}, 
the magnitude range is $9 \le I_{IC} \le 12$, where $I_{IC}$ is the $I$
magnitude of input catalogue. It equals the SuperCosmos photographic $I$
magnitude for faint stars ($11 < I_{IC} < 12$), while for bright stars
($I_{IC} < 11$) it was derived from the Tycho-2 $B_T$ and $V_T$ magnitudes. 
                                                                    
DR2 contains 51,829 radial velocity measurements for 49,327 individual
stars, with a typical radial velocity error of 1.3 to 1.7~\kms\ for data
collected after DR1. Figure~\ref{fig1} shows 
the dependence of the radial velocity errors on temperature, as determined
from RAVE spectra. Radial velocities are marginally better determined for
cool giants than for medium temperature main sequence stars. The errors
increase for stars of an early spectral type. 

DR2 doubles the number of radial velocities previously published in DR1. The
RAVE collaboration is continuing its calibration campaign. The comparison to
external data shows a standard deviation of 1.3~\kms\ for the radial
velocities, which is twice better than for the first data release.          

In addition to radial velocities, the catalogue contains atmospheric
parameters, $\logg$, $\MH$ and $\teff$, for the first time. These parameters
were determined based on 22,407 spectra corresponding to 21,121 individual
stars. A conservative error estimates for these parameters for the average
signal--to--noise ratio of the survey (S/N $\sim 40$) is 400~K for the
effective temperature, 0.5~dex for the logarithm of the gravity ($\logg$) 
and 0.2~dex for the metallicity. These errors depend strongly on S/N. Stars
at the extreme ends of the S/N range have errors $\sim 2$~times
better/worse. The calibration of the RAVE atmospheric parameters is obtained
using a devoted validation campaign where high-resolution spectra of
standard stars have been taken with various instruments and compared to
spectra obtained by RAVE. 

Figures \ref{fig2} to \ref{fig4} show standard deviations of temperature,
gravity, and metallicity as obtained from  repeated observations where a
certain star was observed more than once by RAVE. The results show that
typical errors are only $\simlt 200$~K in effective temperature, $\sim
0.25$~dex in gravity and $\simlt 0.15$~dex in metallicity. 
     
The catalogue is available from the RAVE website www.rave-survey.org or at
the CDS using the VizieR database.

\subsection{Pilot Survey}

The RAVE pilot survey (data collected between April 2003 and February 2006)
is now completed. A data release is scheduled for mid-2009. It contains
about 85,000 radial velocity measurements as well as measurements of stellar
atmospheric parameters for a large fraction of the sample. The release of
the pilot survey will mark a major step forward in the RAVE project, and
subsequent data releases will be based on a new input catalogue drawn from
the DENIS survey.

\subsection{Current Status}

RAVE currently operates at its full potential, observing is scheduled for 25
nights per lunation. The project entered its main survey phase in March
2006. The main survey relies on a new input catalogue based on DENIS I-band
magnitudes. RAVE has so far collected approximately 215,000 spectra in its
main phase, which adds to a total of 301,000 spectra for 270,000 stars when
combining the two phases of the project. Figure~\ref{fig5} plots the density
of spectra per square degree observed before 6th of July 2008. The survey
now covers almost the entire Southern hemisphere with the exception of the
Galactic plane where only a few test observations have been obtained. 

\subsection{Ongoing Efforts}

RAVE's primary goal is to measure accurate radial velocities for stars in
the Southern hemisphere. The quality of the RAVE spectra also enables us to
measure atmospheric parameters ($\logg$, $\MH$ and $\teff$ ) which can be
used to select subsample of the catalogue with specific properties. For
example, the combination of colours and $\logg$ measurements from RAVE
permits an accurate selection of red clump giants \cite[(Veltz et al.
2008]{veltz_2008}, \cite[Siebert et al. 2008)]{siebert_2008}. For this
population, the narrow luminosity function enables us to obtain distances to
20\%\ from the apparent magnitude alone.

However, the measurement of the atmospheric parameters is non-trivial at 
RAVE resolution and the transformation from the measured parameters to the
true parameters relies on calibration data. The RAVE collaboration
constantly acquires data for standard and pseudo-standards stars, using both
the 6dF instrument and high-resolution spectra from other instruments, with
the aim to refine our calibration. 

Despite the medium resolution of the RAVE spectra, chemical abundances for
about 12 elements can be measured with a good accuracy ($\sim 0.2$~dex) in
the high signal to noise spectra ($S/N>80$). The RAVE collaboration puts a
particular effort on measuring and validating these abundances. For this
purpose, RAVE also acquires spectra for nearby stars for which accurate
abundances have been measured using high-resolution spectroscopy. So far,
abundances have been derived for more than 20,000 RAVE targets, and the
resulting measurements will be published in a seperate catalogue.  
Furthermore, we plan to publish a list of fast rotating (and generally hot)
stars where the rotational broadening can be detected at the resolving power
of RAVE spectra.

The knowledge of the 6D phase space requires that we can transform the
proper motions into spatial velocities. This    step relies on good distance
estimates  for individual stars, which can be obtained from comparison of
apparent and absolute magnitudes, the latter inferred from values of
atmospheric parameters. The availability of distances in the near future
will allow to exploit the full potential of the RAVE      catalogue,
permitting to study the detailed shape of the   phase space and reveal new
details of the formation of the Solar neighbourhood.         
                                                                                                                                                            \section{Conclusions}
   
The RAVE collaboration has released a second catalogue in July 2008,
reporting radial velocities for 51,829 spectra and 49,327 different stars,
randomly selected in the magnitude range of $9 < I < 12$ and located more
than $25^\mathrm{o}$ away from the Galactic plane. This release doubles the
size of the previously published catalogue. The typical error of the
published radial velocities is between 1.3 and 1.7~\kms.                    
 
In addition to radial velocities, the catalogue contains, for the first
time, atmospheric parameter measurements for more than 20,000 spectra.
Uncertainties for a typical RAVE star are of the order of 400~K in
temperature, of 0.5~dex in gravity, and of 0.2~dex in metallicity, but the
error depends on the S/N and can vary by a factor of $\sim 2$ for stars at
the extreme ends of the S/N range. Comparison of parameter measurements for 
repeated observations of the same targets indicates that these estimates are
conservative and that the true errors may be smaller. 

The survey continues to collect spectra. Currently more than 300,000 spectra
have been collected and are currently being processed. Acquisition of new
calibration data is also underway, enabling us to refine the atmospheric
parameter measurements as well as radial velocities, and represents an
import ongoing effort. Other current activities include the measurement of
chemical abundances and distances which will be released as companion
catalogues in the future.
 
The RAVE catalogue can be used to study the formation of the Milky Way and,
for example, the collaboration succeeded in refining significantly the
measurement of the local escape velocity using the high velocity stars found
in the RAVE catalogue. Further studies carried out by the collaboration
include the search for vertical stream of matter in the Solar neighborhood;
a new determination of the vertical structure of the Galactic disc; the
study of diffuse interstellar bands in the Galactic plane and their
correlation to the interstellar extinction; the measurement of the
inclination of the velocity ellipsoid towards the Galactic plane at 1~kpc
below the plane.

As the survey progresses and new data become available, the RAVE catalogue
will enable us to refine our view on the structure and formation of the
Milky Way, paving the way for new insights on the formation of
galaxies.\bigskip

Acknowledgements. Funding for RAVE has been provided by the Anglo-Australian
Observatory, the Astrophysical Institute Potsdam, the Australian Research
Council, the German Research foundation, the National Institute for
Astrophysics at Padova, The Johns-Hopkins University, the Netherlands
Research School for Astronomy, the Natural Sciences and Engineering Research
Council of Canada, the Slovenian Research Agency, the Swiss National Science
Foundation, the National Science Foundation of the USA (AST-0508996), the
Netherlands Organisation for Scientific Research, the Particle Physics and
Astronomy Research Council of the UK, Opticon, Strasbourg Observatory, and
the Universities of Basel, Cambridge, and Groningen. The RAVE Web site is at
www.rave-survey.org.

\end{document}